\documentclass[
    tightenlines,
    onecolumn,
    preprintnumbers,
    amsmath,
    amssymb,
    nofootinbib,
    prd,
    showpacs
]{revtex4}
\usepackage[dvips]{graphicx}
\usepackage{color}
\usepackage{amsmath}
\usepackage{amssymb}
\usepackage{bm}
\usepackage{enumerate}
\usepackage{color}
\usepackage{psfrag}
\usepackage{epsfig}
\usepackage{wrapfig}

\newcommand{\beq}{\begin{eqnarray}}
\newcommand{\eeq}{\end{eqnarray}}
\newcommand{\bea}{\begin{eqnarray}}
\newcommand{\eea}{\end{eqnarray}}

\newcommand{\bk}{{\bf k}}

\newcommand{\cI}{{\cal I}}
\newcommand{\cJ}{{\cal J}}

\begin{document}

\title{
Higher order spectra from an initially anisotropic universe
}

\author{Masato Minamitsuji}
\affiliation{Yukawa Institute for Theoretical Physics,
Kyoto University, Kyoto, 606-8502, Japan}

\date{\today}%
\bigskip
\begin{abstract}

In this paper,
we present
the higher order spectra of a scalar field
produced through the 
higher derivative interactions
in the initially anisotropic universe.
Although 
we ignore the backreaction of the scalar field on the geometry,
our analysis should have
much overlap with the quantum fluctuations of the inflaton field
in the anisotropic universe.
We also include
the planar modes whose momenta 
are along the plane which is perpendicular to
the primordial preferred direction,
for which effects of the initial anisotropy
are not suppressed.
The presence of a negative frequency mode
produces features distinguishable  
from the case of the de Sitter inflation.
We also show that richer features appear in the trispectra
due to the primordial anisotropy.
\end{abstract}
\maketitle

\section{Introduction}

Recent measurements by the WMAP satellite
\cite{komatsu,wmap5} 
have suggested
that the observed map of CMB anisotropy
is almost consistent
with the Gaussian and statistically isotropic
primordial fluctuations from inflation.
However,
the issues on
several anomalies
in the recent data of the large-angle CMB temperature map
have been controversial 
(see, e.g., \cite{review,anomaly2,Picon,gawe}).
These observational facts have 
motivated us 
to investigate
the possibilities
of the preinflationary anisotropy
\cite{tpu,gcp,gkp,km,km2,day}.
The cosmic no-hair theorem ensures
that, in the presence of a positive cosmological constant,
an initially anisotropic universe exponentially
approaches the de Sitter spacetime
under the strong or dominant energy condition \cite{wald}.
Therefore, 
the initial universe may be highly anisotropic,
and
its consequences may be inherited 
in the cosmic observables today, such as CMB anisotropy,
if the number of the e-foldings for inflation takes a
minimal value \cite{km,tpu,gcp,gkp}. 
In an expanding Kasner universe,
one of two polarizations is coupled with the scalar mode,
while
the other is decoupled,
leading to an asymmetry between them.
There could be a difference between the amplitudes of 
gravitational waves due to the primordial anisotropy \cite{km2}.

In this paper,
first, we will investigate
the bispectra of
a scalar field
in the initially anisotropic universe
approaching the de Sitter spacetime
(see, e.g., \cite{chen,koyama} for reviews
in the case of the standard inflation).
Amoung two planar branches of the expanding
Kasner spacetime,
the adiabatic vacuum can be well defined 
in the branch where
the initial spacetime structure 
is 
the product of
the two-dimensional Milne universe
and two-dimensional Euclidean space.
Thus,
the inflaton fluctuations are quantized
 in an adiabatic vacuum state
different from the Bunch-Davies vacuum \cite{bd},
which could give a signature distinguishable from
the primordial anisotropy.
Another important effect
from the primordial anisotropy
is the mixing of the scalar and tensor metric perturbations
in terms of the three-dimensional symmetry \cite{km,tpu,gcp,gkp}.
In particular, 
in the nonlinear perturbations such effects may be more significant.
The purpose of this paper is to clarify
the effects of the change of the initial vacuum state
onto the higher order spectra
in a single
scalar field theory \cite{km},
and hence we will not discuss the coupling effects
in the metric perturbations. 
Although 
we ignore the backreaction onto the geometry,
our analysis should have much overlap 
with the cosmological perturbations.
We can expect that
some features appear
in the limit $k_1+ k_2 \sim k_3$,
which is expected to be sensitive to the mixture
of a negative frequency mode \cite{ht}.
The recent work Ref. \cite{day}
has confirmed
part of the above expectations.
The work focused on the nonplanar
high-momentum modes
whose momentum vectors
are not along the plane
where
the effects of the negative frequency mode
are suppressed 
due to the adiabaticity parameter \cite{km,km2}.
We expect that the nonnegligible contributions
of the primordial anisotropy
to the bispectrum would appear 
more significantly in the planar modes
whose momenta are along the plane.
Thus, our analysis includes them.
We will also investigate
the trispectra
in which
the contributions from the primordial anisotropy
may become important even for the nonplanar modes.

The construction of this paper is as follows:
In Sec. 2,
we review the framework
to analyze the higher order spectra
in the anisotropic universe.
In Sec. 3, 
we introduce
the Kasner-de Sitter spacetime
which exhibits an isotropization
due to the presence of a positive cosmological constant.
We also review the 
quantization of a free scalar field in this spacetime.
In Sec. 4, we compute 
the bispectra in the presence
of the higher order time-derivative interactions.
Similarly, we investigate the trispectra in Sec. 5.
In Sec. 6, we shall close this paper after giving a 
brief summary.

\section{Scalar field interactions in the anisotropic universe}

\subsection{Canonical formulation}

In this section, 
we present the framework to analyze the higher order spectra
of a scalar field
in the anisotropic universe,
based on the so-called in-in formalism \cite{weinberg}.

We consider a 
scalar field 
in a curved spacetime background
\bea
\label{saction}
S_\phi
=\int d^4 x\sqrt{-g}{\cal L}
=-\int d^4x\sqrt{-g}
\Big[\frac{1}{2}\big(\partial\phi\big)^2+V\Big],
\eea
where
$V$
represents
the interaction terms
which are of
higher order 
in terms of 
the scalar field and its derivatives.
The form of $V$ will be specified in Sec. III. C.
As the background spacetime,
we consider an anisotropic universe
with the planar symmetry
\bea
ds^2&=&-d\tau^2
+ a(\tau)^2dx^2
+b(\tau)^2
  \big(dy^2+dz^2\big)
=-e^{6\alpha(t)}dt^2
+a(t)^2dx^2
+b(t)^2  \big(dy^2+dz^2\big),
\label{metric}
\eea
where $e^{\alpha}=(ab^2)^{\frac{1}{3}}$
is the avegaged scale factor
and 
$dt=d\tau/e^{3\alpha}$.
We call 
the $(y,z)$-plane the {\it planar} direction,
and simply denote the components of any vector in this
direction by the symbol ``$\perp$.''
We assume
the universe approaches
an isotropic one, $a\to b$,
in the late time limit.
Then
the action of the scalar field Eq. (\ref{saction})
reduces to 
\bea
S_{\phi}&=&-\frac{1}{2}
\int dt\int d^3x e^{6\alpha}
\Big[
-e^{-6\alpha}\dot{\phi}^2
+\frac{1}{a^2}
\phi_{,x}^2
+\frac{1}{b^2}
\phi_{,\perp}^2
+2V\Big],
\eea
where we have defined 
the short-hand notation
as
$\dot{\phi}:=\phi_{,t}$
and 
$\phi_{,\perp}^2:=\phi_{,y}^2+\phi_{,z}^2$.
We use the coordinate $t$ 
to label time
rather than the physical time $\tau$.
The conjugate momentum to the scalar field is given by 
$\pi_{\phi}=\frac{\delta S_{\phi}}{\delta \dot{\phi}}=\dot{\phi}$.
The scalar field and conjugate momentum satisfy
the equal time commutation relations
$\big[\phi(t,x^i),\pi_{\phi}(t,y^i)\big]
=i\delta^{(3)}(x^i-y^i)$ 
and 
$\big[\phi(t,x^i),\phi(t,y^i)\big]
=
\big[\pi_\phi(t,x^i),\pi_\phi(t,y^i)\big]
=0$.
The Hamiltonian density is defined by 
\bea
{\cal H}&=& \dot{\phi}\pi_\phi-{\cal L}
=
\frac{e^{6\alpha}}{2}
\Big( 
 e^{-6\alpha}
 \pi_\phi^2
+ \frac{1}{a^2}\phi_{,x}^2
+ \frac{1}{b^2}
\phi_{,\perp}^2
+2V
\Big).
\eea
The Hamiltonian is given by 
$H\big[\phi(t),\pi_\phi(t);t\big]:=\int d^3 x{\cal H}$.
The equation of motion for the scalar field  is given by 
\bea
\dot{\phi}= i \big[H\big[\phi(t),\pi_\phi(t);t\big],\phi\big],
\quad 
\dot{\pi}_\phi
=i \big[H\big[\phi(t),\pi_\phi(t);t\big],\pi_\phi\big].
\eea
The solutions at $t$
can be expressed in terms of 
a similar operator 
at a very early time $t_0$
through the unitality transformation
\bea
&&\phi(t)
=U^{-1}\big(t,t_0\big)\phi(t_0)
 U\big(t,t_0\big),
\quad
\pi_\phi(t)
=U^{-1}(t,t_0)\pi_\phi(t_0)
 U(t,t_0),
\eea
where $U(t,t_0)$ obeys the differential equation
\bea
\frac{d}{dt}U\big(t,t_0\big)
=-i H\big[\phi(t),\pi_\phi(t);t\big]
   U(t,t_0).
\eea
The initial condition is given by $U(t_0,t_0)=1$.
Let us briefly discuss the choice of the initial time $t_0$.
In cosmology,
the quantization of a scalar field is usually performed
when the corresponding mode is well inside the horizon,
and it is reasonable to choose
$t_0\to -\infty$,
when the effects of the cosmic expansion are
completely
negligible.
But
in any model 
where
the background spacetime 
is modified from the standard one,
the ordinary description of the mode functions
becomes valid 
only after some critical time \cite{ht,day}. 
In the case of an initially anisotropic universe,
we will choose $t_0$
to be the time when
the universe is isotropized, and 
the effective theory description becomes 
valid.
We will discuss this in more detail 
after specifying the interaction terms.

\subsection{Interaction picture}

In calculating $U(t,t_0)$, we decompose ${H}$
into the (quadratic) kinetic part $H_0$
and the interaction part $H_I$
that are of higher order
in the scalar field amplitudes,
\bea
H\big[\phi(t),\pi_\phi(t);t\big]
=H_0\big[\phi(t),\pi_\phi(t);t\big]
+H_I\big[\phi(t),\pi_\phi(t);t\big],
\eea 
where
\bea
H_0\big[\phi(t),\pi_\phi(t);t\big]
&:=&\frac{e^{6\alpha}}{2}
\int d^3 x
\Big(
 e^{-6\alpha}\pi_\phi^2
+
\frac{1}{a^2}
\phi_{,x}^2
+
\frac{1}{b^2}
\phi_{,\perp}^2
\Big),
\nonumber\\
H_I\big[\phi(t),\pi_\phi(t);t\big]
&:=&
e^{6\alpha}
\int d^3 x V.
\label{king}
\eea
We will calculate $U$ as the power series in $H_I$.
We introduce the interaction picture
and define the 
interacting operators $\phi^I(t)$
and $\pi_\phi^I(t)$
whose dependence is determined by the quadratic
part of the Hamiltonian $H_0$,
\bea
\label{int_int}
\dot{\phi}^I(t)=
 i \Big[H_0\big[\phi^I(t),\pi_\phi^I(t);t\big],\phi^I(t)\Big],
\quad
\dot{\pi}_\phi^I(t) =i \Big[H_0\big[\phi^I(t),\pi_\phi^I(t);t\big],\pi_\phi^I(t)\Big],
\eea
where
the initial conditions are given by
$\phi^I(t_0)= \phi(t_0)$
and 
$\pi_\phi^I(t_0)= \pi_\phi(t_0)$.

In evaluating $H_0[\phi^I(t), \pi_\phi^I (t);t]$
we can take the time argument of $\phi^I$ and $\pi_\phi^I$
to be any value,
and we can take $t_0$
so that 
\bea
H_0[\phi^I(t), \pi_\phi^I (t);t]
=H_0[\phi(t_0), \pi_\phi (t_0);t].
\label{16a}
\eea
The solution to Eq. (\ref{int_int}) 
can be written in terms of the unitality transoformation 
$\phi^I(t)= U_0^{-1}(t,t_0) \phi(t_0) U_0(t,t_0)$ 
and
$\pi_\phi^I(t)= U_0^{-1}(t,t_0)\pi_\phi (t_0)U_0 (t,t_0)$,
where $U_0(t,t_0)$ is the solution to the equation
\bea
 \frac{d}{dt} U_0(t,t_0)
=-i H_0
\big[
\phi(t_0),
\pi_\phi(t_0);t
\big]
U_0(t,t_0).
\label{16}
\eea
We obtain
\bea
\frac{d}{dt}
\Big[
 U_0^{-1}(t,t_0)
 U(t,t_0)
\Big]
=-i 
U_0^{-1}(t,t_0)
H_I \big[\phi(t_0),\pi_\phi(t_0);t\big]
U(t,t_0),
\eea
which gives
\bea U(t,t_0)
=U_0(t,t_0){\cal F}(t,t_0),\quad
\frac{d}{dt}{\cal F}(t,t_0)
=-i H_{I}(t) {\cal F}(t,t_0),
\quad
{\cal F}(t_0,t_0)=1.
\eea
We define the shorthand notation for the interaction Hamiltonian
\bea
 H_I (t)
:=U_0 (t,t_0) 
  H_I\big[\phi(t_0),\pi_\phi(t_0);t\big]
  U_0^{-1}(t,t_0)
=H_I \big[
     \phi^I (t),\pi_\phi^I (t);t
     \big].
\eea
The solution for ${\cal F}(t,t_0)$
is given by
\bea
{\cal F}(t,t_0)= T{\rm exp}
 \Big(
-i\int^t_{t_0}
 H_I(t)dt
\Big),
\eea
where $T$ is the time-ordering operator.
The expectation value for some operator ${\cal A}(t)$ is given by 
\bea
\big\langle
{\cal A}(t)
\big\rangle
=\Big\langle
\Big[
{\tilde T} {\rm exp}\Big(i\int_{t_0}^t H_I(t) dt \Big)
\Big]
{\cal A}^I (t)
\Big[
{T} {\rm exp}\Big(-i\int_{t_0}^t H_I(t) dt \Big)
\Big]
\Big\rangle,\label{20}
\eea
where ${\cal A}(t)$ 
can be any product of $\phi (t)$s and $\pi_\phi (t)$s.
For the evaluations of the higher order spectra,
it is more useful to rewrite Eq. (\ref{20}) as
\bea
\big\langle
{\cal A}(t)
\big\rangle
=
\sum_{N=0}^{\infty}
i^N 
\int^{t}_{t_0}dt_N
\int^{t_N}_{t_0} dt_{N-1}
\cdots
\int^{t_2}_{t_0}
dt_1
\Big\langle
\Big[
H_I(t_1),
\Big[
H_I(t_2),\cdots
\Big[
H_I (t_N),{\cal A}^I(t)
\Big]
\cdots
\Big]
\Big]
\Big\rangle
\label{21}.
\nonumber\\
\eea

\section{Preinflationary anisotropic universe and 
the behavior of a scalar field}

In this section, we present the background geometry
and the way to quantize a scalar field in this background.

\subsection{Planar Kasner-de Sitter spacetime}

We consider the Einstein gravity with a positive cosmological constant
$\Lambda>0$
\bea
S_g=\frac{1}{2\kappa^2}\int d^4 x\sqrt{-g}\Big(R-2\Lambda\Big).
\eea
There is the planar Kasner-de Sitter solution
with a two-dimensional
Euclidean symmetry
whose metric is given by Eq. (\ref{metric})
with
\bea \label{yoko2}
a=\sinh ^{
\frac{1}{3}}(3H_0 \tau)
 \tanh^{\frac{2}{3}} \Big(\frac32 H_0 \tau\Big),\quad
b= \sinh ^{
\frac{1}{3}}(3H_0\tau) \coth^{\frac{1}{3}}
\Big(\frac{3}{2}H_0 \tau\Big),
\eea
where $H_0=\sqrt{\frac{\Lambda}{3}}$
is the Hubble constant in the isotropized limit.
The averaged scale factor is given by
\bea
e^{\alpha}:=(ab^2)^{\frac{1}{3}}
=\big(\sinh (3H_0\tau)\big)^{\frac{1}{3}}.
\eea
The spacetime is completely regular everywhere even 
in the limit of the intial time $\tau\to 0$,
in contrast to the other branches of the Kasner spacetime.
The Hubble parameter in each direction is given by
\bea
H_a:=\frac{\dot{a}}{a}
= \frac{H_0}{\sinh(3H_0 \tau)} 
\big( 2+\cosh(3 H_0 \tau) \big),
\quad
H_b:=\frac{\dot{b}}{b}
=H_0 \tanh\big(\frac{3}{2}H_0 \tau\big).
\eea
In this solution,
at the initial times
the spacetime structure 
becomes the product of
a (Milne) patch of the two-dimensional Minkowski spacetime
and the two-dimensional Euclidean space.
Thus it is possible to define the 
initial adianatic vacuum state in this solution.
At the later times for $\tau \to \infty$,
the universe approaches 
the de Sitter spacetime with the expansion $H_0$.
In the other Kasner-de Sitter solutions
which contain initial curvature singularities,
the adiabatic vacuum
cannot be well-defined in the asymptotic past
\cite{km,km2}.

\subsection{Scalar field in the Kasner-de Sitter spacetime}

Here as in Sec. II, we will work in the time coordinate 
$dt=\frac{d\tau}{e^{3\alpha}}$.
Note that these two time coordinates are related
via the simple analytic relation
\bea
\sinh (-3H_0 t)=\frac{1}{\sinh(3H_0\tau)}.
\eea
In the interaction picture,
\bea
\label{ini}
\phi^I
=\int \frac{d^3k}{(2\pi)^{\frac{3}{2}}} 
e^{i{\bf k}{\bf x}}
\phi(t,\bk)
=\int \frac{d^3k}{(2\pi)^{\frac{3}{2}}} 
e^{i{\bf k}{\bf x}}
\big(
 u_{\bf k}(t) a_{\bf k}
+u_{-\bf k}^{\ast}(t)a_{-\bf k}^{\dagger}
\big).
\eea
In our Kasner-de Sitter background,
\bea
 \label{eom}
\Big(
\frac{d^2}{dt^2}
+\Omega(t)^2
\Big)u_{\bf k}(t)
=0\,,
\quad
\Omega^2(t)
:=\frac{2^{4/3}\big(k_\perp^2e^{6H t}+k_1^2\big)}
      {(1-e^{6H t})^{4/3}}=\frac{2^{4/3} k^2}{
\xi^{4/3}(t)}(1- r_\perp^2 \xi(t)),
\eea
where
$k^2=k_1^2+k_{\perp}^2$,
$r_\perp:=k_\perp/k$ and 
\bea
\xi(t) :=1- e^{6Ht}= e^{-6\alpha}(\sqrt{e^{6\alpha}+1}-1),
\eea
varies from one to zero as time $t$ increases from negative infinity to zero.
The late time solution is given by 
\bea
\label{nike}
u_{\bf k}(t)
=A_{\bf k}^{+} u^{(0)}_{\bf k} (t)
+A_{\bf k}^{-} u^{(0)\ast}_{\bf k}(t),
\quad
u^{(0)}_{\bf k} (t)
=\frac{e^{i\frac{k}{H_0}(-3H_0t)^{\frac{1}{3}}}}{\sqrt{2k}}
 \Big[
 \big(-3H_0 t\big)^{\frac{1}{3}}
+\frac{i H_0}{k}
 \Big].
\eea
The coefficients $A_{\bk}^+$ and $A_{\bk}^-$
are determined through the matching of the mode functions.
In Refs. \cite{km,km2},
we have classified modes into
the nonplanar high-momentum modes
and the planar ones.

\vspace{0.2cm}

{\it (1) Nonplanar, high-momentum modes}

\vspace{0.2cm}

The nonplanar high-momentum modes
satisfy
$H_0\ll k_1\sim k_{\perp}$.
For these modes, 
the WKB solutions for the early time
can be directly matched to
mode functions 
in the de Sitter spacetime
at the later times.
The coefficients can be found as
\bea
A_{\bf k}^{+}
=\Big(1-\frac{H_0}{2k}\Big)
\Big(1-i\sqrt{\frac{H_0}{k}}\Big),\quad
A_{\bf k}^{-}
=\frac{1}{2}\Big(\frac{H_0}{k}\Big)^{\frac{3}{2}}
 Q(r_\perp),
\label{coef_np}
\eea
where
$Q(r_\perp):=\frac{2}{3}-r_{\perp}^2$
contains the direction dependence
appearing only in
the negative frequency mode.
$A_{\bk}^{(-)}$
is of order $\epsilon_{\ast}^3$
where $\epsilon_{\ast}$ is the adiabaticity parameter
$\epsilon:=\frac{(\Omega^2)_{,t}}{2\Omega^{3}}$
at the matching time $t=t_{\ast}$.
Hence, 
we expect that these modes
would not give significant effects on 
spectra and cosmic observables at the level
of the power spectrum.

\vspace{0.2cm}

{\it (2) Planar modes}

\vspace{0.2cm}

We then consider the planar modes
with $H_0< k_1\ll k_\perp$,
which are along the 
constant $x$ plane
where there is a two-dimensional
rotational symmetry.
The main difference
from the previous case
is that
there is a temporal violation of
the WKB approximation
during which
effects of the anisotropy are 
encoded into 
the mode functions.
According to Refs \cite{km,km2},
\bea
A_{\bf k}^{+}
=\frac{1}{(1-e^{-2\pi q_1})^{\frac{1}{2}}}
e^{-i\sqrt{\frac{k}{H_0}}
-i\Phi+i(\frac{\pi}{4}-\sqrt{\frac{k_1}{H_0}})},
\quad
A_{\bf k}^{-}
=\frac{e^{-\pi q_1}}{(1-e^{-2\pi q_1})^{\frac{1}{2}}}
e^{i\sqrt{\frac{k}{H_0}}+i\Phi-i(\frac{\pi}{4}
-\sqrt{\frac{k_1}{H_0}})},
\label{coef_p}
\eea
where 
$q_1:=\frac{2^{\frac{2}{3}}k_1}{3H_0}$
and 
$\Phi:=\frac{\sqrt{\pi}\Gamma\big(\frac{1}{3}\big)k}
          {3\times 2^{\frac{1}{3}}\Gamma\big(\frac{5}{6}\big)H_0}
+ O\Big(
   \sqrt{\frac{k}{H_0}}
  \Big)$.
Since $e^{-\pi q_{1}}$ 
may be close to unity, 
$|A_{\bk}^{(-)}|\lesssim |A_{\bk}^{(+)}|$.

To discuss the late time behavior of the mode
functions, it is convenient to work
by choosing
the conformal time
$d\eta=e^{-\alpha}d\tau=e^{2\alpha}dt$,
in which 
the late time mode functions can be rewritten as
\bea
u_{\bf k}(\eta)
=A_{\bf k}^{+} u^{(0)}_{\bf k} (\eta)
+A_{\bf k}^{-} u^{(0)\ast}_{\bf k}(\eta),
\quad
u^{(0)}_{\bf k} (\eta)
=\frac{H_0 e^{-ik \eta}}{\sqrt{2k}}
 \Big(
 -\eta
+\frac{i}{k}
 \Big).
\eea
In the conformal time coordinate,
the averaged scale factor is given by
$e^{\alpha}=-\frac{1}{H_0\eta}$.


\subsection{Interaction Hamiltonian}

In this subsection,
we focus on the choice of $V$ in Eq. (\ref{saction}).
We consider the high energy corrections to the kinetic terms.
Although
in general 
as such corrections
there would be both the time- and 
spatial-derivative terms,
for simplicity
we focus on the time-derivative terms.
The time derivatives of the scalar field in
the Lagrangian density except for the volume factor
can be expanded
in terms of the amplitude 
\bea
\frac{1}{2}\phi_{,\tau}^2
+\sum_{n=3}
  q_n \frac{\phi_{,\tau}^{n}}{M^{2(n-2)}},
\label{kinlag}
\eea
where $q_n$ ($n=3,4,\cdots$) 
are dimensionless constants
which are typically of order unity,
and $M$ is the energy scale
at which the higher order corrections 
become important.
It is also appropriate
to use derivatives 
with respect to the proper time $\tau$,
rather than another time coordinate, such as $t$.
We may regard our scalar field theory
as the low energy limit of a complete UV theory.
appear just as low energy corrections.
In the interaction picture,
the leading order quadratic term gives
the ordinary kinetic term,
and higher order ones give interactions.
Since we are particularly interested in the three- and four-point interactions,
we focus on the terms of $n=3$ and $n=4$.
We expect that 
the time-derivative interactions
would give us information 
enough
to see at least the difference of the anisotropic background
from the de Sitter inflation.

Thus
we choose $V$ in Eq. (\ref{saction}) as
\bea
V
=
e^{-6\alpha}
\pi_{\phi}^2
\Big[
q_3 e^{-3\alpha}\frac{\pi_{\phi}}{M^2}
+q_4 e^{-6\alpha}
\Big(\frac{\pi_{\phi}}{M^2}\Big)^2
\Big].
\label{33}
\eea
Using
$\pi_{\phi}=\dot{\phi}$ and 
$dt=e^{-3\alpha}d\tau$, 
Eq. (\ref{33}) is rewritten as 
\bea
V= \phi_{,\tau}^2
\Big[
  q_3 \frac{\phi_{,\tau}}{M^2}
 +q_4 \Big(\frac{\phi_{,\tau}}{M^2}\Big)^2
\Big],
\eea
which agrees with the $n=3$ and $n=4$ terms in
Eq. (\ref{kinlag}).
From Eq. (\ref{king}),
the interaction Hamiltonian is given by
\bea
H_I\big[\phi(t),\pi(t);t\big]
&=&
H_I^{(3)}\big[\phi(t),\pi(t);t\big]
+H_I^{(4)}\big[\phi(t),\pi(t);t\big]
\nonumber\\
&=&
e^{6\alpha}
\int d^3 x
\phi_{,\tau}^2
\Big[
 q_3\frac{\phi_{,\tau}}{M^2} 
+q_4 \Big(\frac{\phi_{,\tau}}{M^2}\Big)^2
\Big],
\nonumber\\
H_I^{(3)}\big[\phi(t),\pi(t);t\big]
&:=&
e^{6\alpha}
\int d^3 x 
 \frac{q_3}{M^2} \phi_{,\tau}^3
=e^{-3\alpha}
\int d^3 x 
 \frac{q_3}{M^2} \dot{\phi}^3,
\nonumber\\
H_I^{(4)}\big[\phi(t),\pi(t);t\big]
&:=&
e^{6\alpha}
\int d^3 x 
 \frac{q_4}{M^4} \phi_{,\tau}^4
=e^{-6\alpha}
\int d^3 x 
 \frac{q_4}{M^4} \dot{\phi}^4.
\label{min}
\eea

Let us focus more on the reasons for considering the derivative interactions. 
The first reason is that in the metric perturbation theory
(most of) the interactions are given 
by the interactions of derivatives \cite{chen}.
The second reason is that
higher derivative interactions
may enhance the higher order spectra \cite{ht,cre}
because the interactions depend on the inverse power
of the (averaged) scale factor,
which is large in the past.
Although 
we ignore the backreaction of the scalar field on the geometry,
our analysis should contain 
overlap with the realistic analysis on the quantum fluctuations.
Since
$\langle |\phi_{,\tau}|^2\rangle \simeq H_0^4$,
the scalar field does not backreact significantly on the background
geometry so long as $H_0< M_p$.
In addition, 
in the equation of motion of the scalar field 
the corrections of the higher order derivative terms 
to the kinetic term 
appear
in order of $\frac{H_0^2}{M^2}$.
Thus, the higher-order derivative interactions
can always be treated as perturbations
so long as $M> H_0$,
and then the early time WKB solutions and 
 late time de Sitter ones given by Eq. (\ref{nike})
are good approximations of the mode functions
with a very high accuracy.
Hence, our perturbative arguments
are valid
when the relation $H_0<M\lesssim M_p$ is satisfied.
We choose the cut-off time for the time integral
to be $\frac{k}{e^{\alpha}}= M$
when the physical momentum for the corresponding mode
becomes below the energy scale
of the derivative interactions.
This gives $-\eta_0=\frac{M}{kH_0}$,
where $k$ is the total momentum for the 
relevant spectra,
i.e. $k=k_1+k_2+k_3$ for the bispectra
and 
$k=k_1+k_2+k_3+k_4$ for the trispectra.
This time becomes later than the time 
when the de Sitter mode function becomes valid,
$-\eta_0<-\eta_\ast:=\frac{1}{\sqrt{k H_0}}$,
leading to $k<\frac{M^2}{H_0}$.
In other words, our interest is
in the energy scale satisfying
\bea
\frac{M}{H_0}<\epsilon_{\ast}^{-1},
\label{sam}
\eea
for a corresponding number of $k$.
Thus, in the subsequent parts of this paper
we will focus on these modes.

In the next sections,
we will present the bispectra and trispectra
following the formulation presented in the previous section.
The leading order contribution to the bispectrum 
is given by the three-point contact interaction diagram.
Similarly, the leading order contribution to the trispectra
is given by two kinds of the connected diagrams.
The first one
is the contact interaction diagram 
[see, e.g., Fig. 1 (A) in \cite{tri}],
and the second one is 
the scalar exchange interaction diagram
[see, e.g., Fig. 1 (B) in \cite{tri}].
We expect that loop corrections are
suppressed 
and ignore them in the subsequent discussions.

\subsection{Power spectra}

Before closing this section,
we show
the power spectra 
including the leading order corrections
due to the primordial anisotropy,
for both the nonplanar high-momentum modes
and the planar modes,
which were obtained in \cite{km,km2}.
Note that, in contrast to the power spectra 
obtained from the standard isotropic scenario,
they depend not only on the
magnitude of the momentum vector $k=|\bk|$,
but also on its direction.
For the nonplanar high-momentum modes,
the power spectrum is given by
\bea
P({\bk})
=\Big(\frac{H_0}{2\pi}\Big)^2
  \Big(
   1
  +Q(r_\perp)\Big(\frac{H_0}{k}\Big)^{3/2}
   \cos\Big(2\sqrt{\frac{k}{H_0}}\Big)
  \Big),
\eea
where $Q(r_{\perp})$ is defined below Eq. (\ref{coef_np}).
Thus the corrections due to the primordial anisotropy
are suppressed
by $\epsilon_{\ast}^3$.

On the other hand,
for the planar modes,
the power spectrum is given by 
\bea
P({\bk})
=\Big(\frac{H_0}{2\pi}\Big)^2
  \Big(
\coth (\pi q_1)
-\frac{\sin(2\Phi)}{\sinh(\pi q_1)}
  \Big),
\eea
where $q_1$ and $\Phi$ are defined below Eq. (\ref{coef_p}).
Thus the corrections due to the primordial anisotropy
are not suppressed.


\section{Bispectra}

The leading order contribution
to the bispectra 
is given
by the contact interaction 
\bea
\langle
\phi(t,{\bf k}_1)\phi(t,\bk_2)\phi(t,\bk_3)
\rangle
=i\int_{t_0}^{t}dt_1
\Big\langle
\big[H_I^{(3)} (t_1),\phi^I(t,\bk_1)\phi^I(t,\bk_2)\phi^I(t,\bk_3)\big]
\Big\rangle.
\eea
The bispectrum in the $t\to 0$ limit takes the form of
\bea
\langle
\phi(0,{\bf k}_1)\phi(0,\bk_2)\phi(0,\bk_3)
\rangle
=(2\pi)^{-3}\frac{q_3}{M^2}B(\bk_1,\bk_2,\bk_3)
\delta(\bk_1+\bk_2+\bk_3).
\eea
After computations,
we obtain
\bea
&&
B(\bk_1,\bk_2,\bk_3)
\nonumber\\
&=&
-\frac{3H_0^5}{4(k_1k_2k_3)}
\Big\{
i\big(A^+_{{\bf k}_1}-A^-_{{\bf k}_1}\big)^{\ast}
\big(A^+_{{\bf k}_2}-A^-_{{\bf k}_2}\big)^{\ast}
\big(A^+_{{\bf k}_3}-A^-_{{\bf k}_3}\big)^{\ast}
\nonumber\\
&\times&
\Big[
-A^+_{\bk_1}A^+_{\bk_2}A^+_{\bk_3}
 \cI(k_1,k_2,k_3)
+A^-_{\bk_1}A^-_{\bk_2}A^-_{\bk_3}
 \cI(-k_1,-k_2,-k_3)
\nonumber\\
&+&
\Big( A^+_{\bk_1}A^+_{\bk_2}A^-_{\bk_3}
 \cI(k_1,k_2,-k_3)
+ {\rm 2 \,\,perms}\Big)
-
\Big( A^+_{\bk_1}A^-_{\bk_2}A^-_{\bk_3}
 \cI(k_1,-k_2,-k_3)
+ {\rm 2 \,\,perms}\Big)
\Big]
\nonumber\\
&+&
 { C. C.}
\Big\},
\eea
where
\bea
&&\cI(p_1,p_2,p_3)
\nonumber\\
&:=&
\frac{1}{(p_1+p_2+p_3)^3}
\Big\{
2i
- e^{-i(p_1+p_2+p_3)\eta_0}
\big[ 
 2i
-2(p_1+p_2+p_3)\eta_0
-i(p_1+p_2+p_3)^2\eta_0^2
\big]
\Big\},
\nonumber\\
\eea
and
$\cI(p_1,p_2,p_3)^{\ast}
= \cI(-p_1,-p_2,-p_3)$.
In the limit of $p_1+p_2+p_3\to 0$,
the function $\cI$ remains finite,
\bea
\cI (p_1,p_2,p_3)
=\frac{\eta_0^3}{3}
+O(p_1+p_2+p_3).
\eea
Since $|\cI(p_1,p_2,p_3)|$
is increasing
in the limit of $p_1+p_2+p_3\to 0$,
the bispectrum
would have peaks 
in the limits of $k_1\to k_2+k_3$,
$k_2\to k_1+k_3$ and $k_3\to k_1+k_2$ (case I).
The bispectrum can also
have a peak in the limits $k_1=k_2\gg k_3$,
$k_2=k_3\gg k_1$, and $k_3=k_1\gg k_2$ (case II).
We discuss each case separately.

\subsubsection{Case I}

In the limit of 
of $k_1\to k_2+k_3$,
the leading order contribution to the bispectrum
is given by
\bea
&&
B(\bk_1,\bk_2,\bk_3)
\nonumber\\
&\to&
\frac{
 H_0^5 \eta_0^3}
     {2(k_1 k_2 k_3) 
}
{\rm Im}
\Big\{
\big(A^+_{{\bf k}_1}-A^-_{{\bf k}_1}\big)^{\ast}
\big(A^+_{{\bf k}_2}-A^-_{{\bf k}_2}\big)^{\ast}
\big(A^+_{{\bf k}_3}-A^-_{{\bf k}_3}\big)^{\ast}
\big(A_{\bk_1}^- A_{\bk_2}^+ A_{\bk_3}^+
-
A_{\bk_1}^+ A_{\bk_2}^- A_{\bk_3}^-\big)
\Big\}.
\nonumber\\
\eea
Similar results are obtained in the limits of
$k_2\to k_1+k_3$ and $k_3\to k_1+k_2$.
For the bispectrum for
the nonplanar high-momentum modes,
\bea
&&{\rm Im}
\Big\{
\big(A^+_{{\bf k}_1}-A^-_{{\bf k}_1}\big)^{\ast}
\big(A^+_{{\bf k}_2}-A^-_{{\bf k}_2}\big)^{\ast}
\big(A^+_{{\bf k}_3}-A^-_{{\bf k}_3}\big)^{\ast}
\big(A_{\bk_1}^- A_{\bk_2}^+ A_{\bk_3}^+
-
A_{\bk_1}^+ A_{\bk_2}^- A_{\bk_3}^-\big)
\Big\}
\simeq
\frac{Q(r_{\perp,1})}{2}
\frac{H_0^2}{k_1^{2}}.
\nonumber\\
&&
\eea
Thus
\bea
&&
B(\bk_1,\bk_2,\bk_3)
\to
\frac{
H_0^7 \eta_0^3}
     {4(k_1^3 k_2 k_3) }
Q(r_{\perp,1}).
\label{bi_lim}
\eea
The ratio of Eq. (\ref{bi_lim}) to 
the case of the de Sitter inflation,
Eq. (\ref{bds}),
is given by 
\bea
Q(r_{\perp,1})H_0^2 k_1\eta_0^3
=Q(r_{\perp,1})
\frac{M^3}{H_0^3}
\frac{k_1H_0^2}{k^3},
\eea 
which is of order $\frac{M^3}{H_0^3}\epsilon_{\ast}^4$.
Thus the direction dependent bispectrum
can be as large as 
that in the de Sitter inflation
for $\frac{M}{H_0}>\epsilon_{\ast}^{-\frac{4}{3}}$.
But this is never satisfied because of Eq. (\ref{sam}).

On the other hand,
for the planar modes,
using the phase factor
$\psi:=-\sqrt{\frac{k}{H_0}}-\Phi
+\big(\frac{\pi}{4}-\sqrt{\frac{k_1}{H_0}}\big)$
the bispectrum in the same limit
is given by 
\bea
&&
B(\bk_1,\bk_2,\bk_3)
\to
-\frac{
    H_0^5 \eta_0^3}
     {2(k_1 k_2 k_3) 
}
\frac{e^{-\pi q_{1,1}}\sin (2\psi_1)}
{(1-e^{-2\pi q_{1,1}})
(1-e^{-2\pi q_{1,2}})
(1-e^{-2\pi q_{1,3}})}.
\label{p_b}
\eea
The ratio to the case of the de Sitter inflation
is given by
\bea
\frac{M^3}{H_0^3}
\frac{k_1^3}{k^3}
\frac{e^{-\pi q_{1,1}}\sin (2\psi_1)}
{(1-e^{-2\pi q_{1,1}})
(1-e^{-2\pi q_{1,2}})
(1-e^{-2\pi q_{1,3}})},
\eea
which is greater than unity
since $\frac{M}{H_0}\gg 1$.
This bound is consistent with Eq. (\ref{sam})
for $1<\frac{M}{H_0}<\epsilon_{\ast}^{-1}$.
Therefore, 
such a bispectrum is
more important than
for the case of the nonplanar high-momentum modes.
This bound can be consistent with Eq. (\ref{sam}).

Similar results can be obtained 
for the limits of $k_2\to k_3+ k_1$, and $k_3\to k_1+k_2$.

\subsubsection{Case II}

In the limit of $k_1= k_2\gg k_3$,
the leading order contribution to the bispectrum
from the nonplanar high-momentum modes
is given by 
\bea
B(\bk_1,\bk_2,\bk_3)
&\to&
\frac{
 H_0^5 \eta_0^3}
     {2(k_1 k_2 k_3) 
}
{\rm Im}
\Big\{
\big(A^+_{{\bf k}_1}-A^-_{{\bf k}_1}\big)^{\ast}
\big(A^+_{{\bf k}_2}-A^-_{{\bf k}_2}\big)^{\ast}
\big|A^+_{{\bf k}_3}-A^-_{{\bf k}_3}\big|^2
\big(A_{\bk_1}^+ A_{\bk_2}^- 
   + A_{\bk_1}^- A_{\bk_2}^+
\big)
\Big\}
\nonumber\\
&\simeq &
\frac{
   H_0^7 \eta_0^3}
     {4k_1^4 k_3 
}
\big(
 Q(r_{\perp,1})
+Q(r_{\perp,2})
\big).
\label{lin}
\eea
The ratio of Eq. (\ref{lin})
to the case of the de Sitter inflation,
Eq. (\ref{bds}),
is given by 
\bea
k_1 H_0^2\eta_0^3
\Big(Q(r_{\perp,1})+Q(r_{\perp,2})\Big)
=\frac{M^3}{H_0^3}
\frac{k_1H_0^2}{k^3}
\Big(Q(r_{\perp,1})+Q(r_{\perp,2})\Big),
\eea
which is of order $\frac{M^3}{H_0^3}\epsilon_{\ast}^4$.
Thus the direction dependent bispectrum
can be as large as that 
in the de Sitter inflation
for $\frac{M}{H_0}>\epsilon_{\ast}^{-\frac{4}{3}}$.
But this is never satisfied because of Eq. (\ref{sam}).
In addition,
assuming that $k=k_1+k_2+k_3$ is fixed,
the ratio of 
the bispectrum of case II, Eq. (\ref{lin}), 
to that of case I, Eq. (\ref{bi_lim})
(with the permutation to $k_3\to k_1+k_2$)
is given by $\frac{k}{k_3}$.
Thus the bispectrum in case II
is enhanced to that in case I
by the ratio $\frac{k}{k_3}\gg 1$.

The leading order contribution
from the planar modes is given by
\bea
&&
B(\bk_1,\bk_2,\bk_3)
\simeq
-\frac{
  H_0^5 \eta_0^3}
     {2(k_1^2 k_3) 
}
\frac{e^{-\pi q_{1,1}}\sin(2\psi_{1})+e^{-\pi q_{1,2}}\sin(2\psi_{2})}
    {(1-e^{-2\pi q_{1,1}})
     (1-e^{-2\pi q_{1,2}})
     (1-e^{-2\pi q_{1,3}})}.
\label{p_c}
\eea
The ratio to the case of the de Sitter inflation
is given by
\bea
\frac{M^3}{H_0^3}
\frac{k_1^3}{k^3}
\frac{e^{-\pi q_{1,1}}\sin (2\psi_1)
+e^{-\pi q_{1,2}}\sin (2\psi_2)}
{(1-e^{-2\pi q_{1,1}})
(1-e^{-2\pi q_{1,2}})
(1-e^{-2\pi q_{1,3}})},
\eea
which is greater than unity
since $\frac{M}{H_0}\gg 1$.
This bound is consistent with Eq. (\ref{sam})
for $1<\frac{M}{H_0}<\epsilon_{\ast}^{-1}$.
The ratio of Eq. (\ref{p_b}) with permutation to $k_3\to k_1+ k_2$
 to Eq. (\ref{p_c})
is given by 
\bea
\frac{k_3}{k}
\frac{e^{-\pi q_{1,3}}\sin (2\psi_3)}
     {e^{-\pi q_{1,1}}\sin (2\psi_1)+e^{-\pi q_{1,2}}\sin (2\psi_2)},
\eea
which becomes smaller than unity
as for the nonplanar high-momentum modes.

A similar result can be obtained for 
the cases of $k_2=k_3\gg k_1$ and $k_3=k_1\gg k_2$.


\subsubsection{Features of the bispectra}

We summarize the features of the bispectra
in our background spacetime
and compare them with
those in the de Sitter inflation.
In our background
the bispectrum in the limit of case I
(for example, for $k_1\to k_2+k_3$ )
is greater
than that in the limit of case II
(for example, for $k_1=k_2\gg k_3$),
by a factor of $\frac{k}{k_3}$,
where $k$ is the typical
magnitude of the momentum vectors.

We then discuss the direction dependences
of the bispectra.
Though we focus on the nonplanar high-momentum modes, 
the essence is the same 
also for the planar modes.
In case I, 
for example, for $k_1\to k_2+k_3$,
the bispectrum
is proportional to $Q(r_{\perp,1})$
and hence 
depends on the direction of the momentum
vector ${\bf k}_1$.
In case II,
for example, for $k_1=k_2\gg k_3$,
it is proportional to $Q(r_{\perp,1})+Q(r_{\perp,2})$,
and hence depends 
on the directions of the momenta ${\bf k}_1$ and ${\bf k}_2$,
which have shorter wavelengths
than ${\bf k}_3$.
Thus the bispectra 
are direction dependent,
and the dependence is significantly
different in various limiting cases.

We finally compare the amplitude of the bispectra
with those in the de Sitter inflation.
For the planar modes,
the bispectra
in both the limits of cases I and II
can be greater than those in
the de Sitter spacetime.
On the other hand,
for the nonplanar high-momentum modes,
the bispectrum in both cases
cannot exceed those
in the de Sitter inflation.
Thus, combined with the direction dependence,
the bispectra for the planar modes
in all the limiting cases
would be particularly interesting
to distinguish models.

In the next section,
we will investigate the trispectra.
In contrast to the bispectra, 
even for the nonplanar high-momentum modes,
the trispectra could be more important than those in the de Sitter
inflation.

\section{Trispectra}

The leading order contributions to the
trispectra are given
by the contact interaction
as welll as the scalar exchange interaction.

\subsection{Contribution from the contact interaction}

The leading order contribution of the four-point interaction
is given by 
\bea
&& \langle \phi(t,{\bf k}_1)\phi(t,\bk_2)
\phi(t,\bk_3)\phi(t,\bk_4)\rangle
=i\int_{t_0}^{t}dt_1
\Big\langle
\big[H_I^{(4)} (t_1),\phi^I(t,\bk_1)\phi^I(t,\bk_2)
\phi^I(t,\bk_3)\phi^I(t,\bk_4)\big]
\Big\rangle.
\eea
The trispectrum in the $t\to 0$ limit takes the following form:
\bea
\langle
\phi(0,{\bf k}_1)\phi(0,\bk_2)\phi(0,\bk_3)\phi(0,\bk_4)
\rangle
=(2\pi)^{-3}\frac{q_4}{M^4}
T_c(\bk_1,\bk_2,\bk_3,\bk_4)
\delta(\bk_1+\bk_2+\bk_3+\bk_4).
\eea
After computations, we obtain
\bea
T_c(\bk_1,\bk_2,\bk_3,\bk_4)
&=&
\frac{6iH_0^8}{(k_1k_2k_3k_4)
}
\Big[
\big(A^+_{{\bf k}_1}-A^-_{{\bf k}_1}\big)^{\ast}
\big(A^+_{{\bf k}_2}-A^-_{{\bf k}_2}\big)^{\ast}
\big(A^+_{{\bf k}_3}-A^-_{{\bf k}_3}\big)^{\ast}
\big(A^+_{{\bf k}_4}-A^-_{{\bf k}_4}\big)^{\ast}
\nonumber\\
&\times&
\Big\{
\Big[
-A^+_{{\bf k}_1}A^+_{{\bf k}_2}
 A^+_{{\bf k}_3}A^+_{{\bf k}_4}
 {\cal J}(k_1,k_2,k_3,k_4)
-A^-_{{\bf k}_1}A^-_{{\bf k}_2}
 A^-_{{\bf k}_3}A^-_{{\bf k}_4}
 {\cal J}(-k_1,-k_2,-k_3,-k_4)
\nonumber\\
&+&\Big(A^+_{{\bf k}_1}A^+_{{\bf k}_2}
   A^+_{{\bf k}_3}A^-_{{\bf k}_4}
  {\cal J}(k_1,k_2,k_3,-k_4)
 +{\rm 3\,\, perms}
\Big)
\nonumber\\
&-&
\Big(A^+_{{\bf k}_1}A^+_{{\bf k}_2}
   A^-_{{\bf k}_3}A^-_{{\bf k}_4}
  {\cal J}(k_1,k_2,-k_3,-k_4)
 +{\rm 5\,\, perms}
\Big)
\nonumber\\
&+&
\Big(A^+_{{\bf k}_1}A^-_{{\bf k}_2}
   A^-_{{\bf k}_3}A^-_{{\bf k}_4}
  {\cal J}(k_1,-k_2,-k_3,-k_4)
 +{\rm 3 \,\, perms}
\Big)
\Big\}
\nonumber\\
&-& C.C.
\Big],
\eea
where
we have defined
\bea
&&\cJ(p_1,p_2,p_3,p_4)
\nonumber\\
&:=&
\frac{1}{4(p_1+p_2+p_3+p_4)^5}
\Big\{
-24i
+e^{-i(p_1+p_2+p_3+p_4)\eta_0}
\Big[
24i
-(p_1+p_2+p_3+p_4)\eta_0
\nonumber\\
&\times&
\Big(
24
-i(p_1+p_2+p_3+p_4)\eta_0
\big(
-12
-(p_1+p_2+p_3+p_4)\eta_0
\nonumber\\
&\times&
\big(
4i
-(p_1+p_2+p_3+p_4)\eta_0
\big)
\big)
\Big)
\Big]
\Big\}.
\eea
In the limit of $p_1+p_2+p_3+p_4\to 0$,
the function $\cJ$ remains finite
as
\bea
\cJ(p_1,p_2,p_3,p_4)
= 
\frac{1}{20}\eta_0^5+ 
O(p_1+p_2+p_3+p_4).
\eea
Thus the trispectrum from the 
contact interaction depends 
on the size of four vectors $k_i$ ($i=1,2,3,4$),
and not on the
size of two independent relative vectors
$k_{12}=|\bk_1+\bk_2|$
and 
$k_{14}=|\bk_1+\bk_4|$,
since there is no internal line in the 
diagram.
Since $|\cJ(p_1,p_2,p_3,p_4)|$
is increasing
in the limit of $p_1+p_2+p_3+p_4\to 0$,
the trispectrum
has the maximum amplitude
in the limits
of $k_1\to k_2+k_3+k_4$,
$k_2\to k_1+k_3+k_4$,
$k_3\to k_1+k_2+k_4$,
and 
$k_4\to k_1+k_2+k_3$ 
(case I),
as well as
$k_1+k_2\to k_3+k_4$,
and 
$k_1+k_4\to k_2+k_3$
(case II).
We also discuss the limits of 
$k_1= k_2\gg k_{12}$ and $k_3= k_4\gg k_{12}$,
and similarly $k_1= k_4\gg k_{14}$ and $k_2= k_3\gg k_{14}$
(case III).

\subsubsection{Case I}

In the limit of $k_1\to k_2+k_3+k_4$,
the leading order contribution to the trispectrum is 
given by
\bea
T_c(\bk_1,\bk_2,\bk_3,\bk_4)
&\to &
-\frac{3
 H_0^8 \eta_0^5}
    {5(k_1k_2k_3k_4)
}
{\rm Im}
\Big\{
\big(A^+_{{\bf k}_1}-A^-_{{\bf k}_1}\big)^{\ast}
\big(A^+_{{\bf k}_2}-A^-_{{\bf k}_2}\big)^{\ast}
\big(A^+_{{\bf k}_3}-A^-_{{\bf k}_3}\big)^{\ast}
\big(A^+_{{\bf k}_4}-A^-_{{\bf k}_4}\big)^{\ast}
\nonumber\\
&\times&
\Big(
A_{\bk_1}^{-}
A_{\bk_2}^{+}
A_{\bk_3}^{+}
A_{\bk_4}^{+}
+
A_{\bk_1}^{+}
A_{\bk_2}^{-}
A_{\bk_3}^{-}
A_{\bk_4}^{-}
\Big)
\Big\}.
\eea
For the trispectrum 
for the nonplanar high-momentum modes,
we obtain
\bea
&&
{\rm Im}
\Big\{
\big(A^+_{{\bf k}_1}-A^-_{{\bf k}_1}\big)^{\ast}
\big(A^+_{{\bf k}_2}-A^-_{{\bf k}_2}\big)^{\ast}
\big(A^+_{{\bf k}_3}-A^-_{{\bf k}_3}\big)^{\ast}
\big(A^+_{{\bf k}_4}-A^-_{{\bf k}_4}\big)^{\ast}
\Big(
A_{\bk_1}^{-}
A_{\bk_2}^{+}
A_{\bk_3}^{+}
A_{\bk_4}^{+}
+
A_{\bk_1}^{+}
A_{\bk_2}^{-}
A_{\bk_3}^{-}
A_{\bk_4}^{-}
\Big)
\Big\}
\nonumber\\
&&\simeq 
\frac{Q(r_{\perp,1})}{2}
\frac{H_0^2}{k_1^2}.
\eea 
Hence
\bea
T_c(\bk_1,\bk_2,\bk_3,\bk_4)
\to 
-\frac{3
H_0^{10} \eta_0^5}
    {10(k_1^3 k_2 k_3 k_4)
     }
Q(r_{\perp, 1}).
\label{con_lim_a}
\eea
The ratio of Eq. (\ref{con_lim_a})
to the case of the de Sitter inflation,
Eq. (\ref{tds1}),
is given by 
\bea
k_1^3 H_0^2\eta_0^5Q(r_{\perp,1})
=\frac{M^5}{H_0^5}
\frac{k_1^3 H_0^2}{k^5}Q(r_{\perp,1}),
\eea
which is of order $\frac{M^5}{H_0^5}\epsilon_{\ast}^4$.
Thus the direction dependent trispectrum 
can be as large as 
that in the de Sitter inflation
for $\frac{M}{H_0}>\epsilon_{\ast}^{-\frac{4}{5}}$.
With the condition for the modes Eq. (\ref{sam}), 
the energy scale for the interaction is bounded as
$\epsilon_{\ast}^{-\frac{4}{5}}<\frac{M}{H_0}<\epsilon_{\ast}^{-1}$.

On the other hand,
for the planar modes,
the trispectrum 
in the same limit
is given by
\bea
T_c(\bk_1,\bk_2,\bk_3,\bk_4)
&\to &
\frac{3
H_0^8 \eta_0^5}
    {5(k_1 k_2 k_3 k_4)
 }
\frac{e^{-\pi q_{1,1}}\sin (2\psi_1)}
{(1-e^{-2\pi q_{1,1}})
(1-e^{-2\pi q_{1,2}})
(1-e^{-2\pi q_{1,3}})
(1-e^{-2\pi q_{1,4}})}.
\label{t_p1}
\eea
The ratio to the case of the de Sitter inflation
is given by
\bea
\frac{M^5}{H_0^5}
\frac{k_1^5}{k^5}
\frac{e^{-\pi q_{1,1}}\sin (2\psi_1)}
{(1-e^{-2\pi q_{1,1}})
(1-e^{-2\pi q_{1,2}})
(1-e^{-2\pi q_{1,3}})
(1-e^{-2\pi q_{1,4}})},
\eea
which is greater than unity
since $\frac{M}{H_0}\gg 1$.
This bound is consistent with Eq. (\ref{sam})
for $1<\frac{M}{H_0}<\epsilon_{\ast}^{-1}$.

Similar results can be obtained for 
the cases of 
$k_2\to k_1+k_3+k_4$, 
$k_3\to k_1+k_2+k_4$,
and 
$k_4\to k_1+k_2+k_3$.

\subsubsection{Case II}

In the limit $k_1+k_2\to k_3+k_4$,
the leading order contribution to the trispectrum is given by
\bea
T_c(\bk_1,\bk_2,\bk_3,\bk_4)
&\to &
\frac{3
H_0^8 \eta_0^5}
    {5(k_1 k_2 k_3 k_4)
}
{\rm Im}
\Big\{
\big(A^+_{{\bf k}_1}-A^-_{{\bf k}_1}\big)^{\ast}
\big(A^+_{{\bf k}_2}-A^-_{{\bf k}_2}\big)^{\ast}
\big(A^+_{{\bf k}_3}-A^-_{{\bf k}_3}\big)^{\ast}
\big(A^+_{{\bf k}_4}-A^-_{{\bf k}_4}\big)^{\ast}
\nonumber\\
&\times&
\Big(
A_{\bk_1}^{-}
A_{\bk_2}^{-}
A_{\bk_3}^{+}
A_{\bk_4}^{+}
+
A_{\bk_1}^{+}
A_{\bk_2}^{+}
A_{\bk_3}^{-}
A_{\bk_4}^{-}
\Big)
\Big\}.
\eea
We see that this type of trispectrum
can be much more important than 
in the previous case.
For the trispectrum
for the nonplanar high-momentum modes,
we obtain
\bea
&&{\rm Im}
\Big\{
\big(A^+_{{\bf k}_1}-A^-_{{\bf k}_1}\big)^{\ast}
\big(A^+_{{\bf k}_2}-A^-_{{\bf k}_2}\big)^{\ast}
\big(A^+_{{\bf k}_3}-A^-_{{\bf k}_3}\big)^{\ast}
\big(A^+_{{\bf k}_4}-A^-_{{\bf k}_4}\big)^{\ast}
\nonumber\\
&\times&
\Big(
A_{\bk_1}^{-}
A_{\bk_2}^{-}
A_{\bk_3}^{+}
A_{\bk_4}^{+}
+
A_{\bk_1}^{+}
A_{\bk_2}^{+}
A_{\bk_3}^{-}
A_{\bk_4}^{-}
\Big)
\Big\}
\nonumber\\
&\simeq &
\frac{Q(r_{\perp,1})Q(r_{\perp,2})}{4}
\frac{H_0^{7/2}}{k_1^{3/2}k_2^{3/2}}
\Big(
 \frac{1}{k_1^{\frac{1}{2}}}
+\frac{1}{k_2^{\frac{1}{2}}}
\Big)
+\frac{Q(r_{\perp,3})Q(r_{\perp,4})}{4}
\frac{H_0^{7/2}}{k_3^{3/2}k_4^{3/2}}
\Big(
 \frac{1}{k_3^{\frac{1}{2}}}
+\frac{1}{k_4^{\frac{1}{2}}}
\Big).
\eea
Thus
\bea
T_c(\bk_1,\bk_2,\bk_3,\bk_4)
&\to &
\frac{3
 H_0^{\frac{23}{2}} \eta_0^5}
    {20(k_1 k_2 k_3 k_4)
}
\Big[
\frac{Q(r_{\perp, 1})Q(r_{\perp,2})}
     {k_1^{\frac{3}{2}}k_2^{\frac{3}{2}}}
\Big(
 \frac{1}{k_1^{\frac{1}{2}}}
+\frac{1}{k_2^{\frac{1}{2}}}
\Big)
+\frac{Q(r_{\perp,3})Q(r_{\perp,4})}
     {k_3^{\frac{3}{2}}k_4^{\frac{3}{2}}}
\Big(
 \frac{1}{k_3^{\frac{1}{2}}}
+\frac{1}{k_4^{\frac{1}{2}}}
\Big)
\Big].
\nonumber\\
\label{con4_lim}
\eea
The ratio of Eq. (\ref{con4_lim})
to the case of the de Sitter inflation
Eq. (\ref{tds1})
is given by 
\bea
&&(k_1+k_2)^5\eta_0^5 H_0^{\frac{7}{2}}
\Big[
\frac{Q(r_{\perp, 1})Q(r_{\perp,2})}
     {k_1^{\frac{3}{2}}k_2^{\frac{3}{2}}}
\Big(
 \frac{1}{k_1^{\frac{1}{2}}}
+\frac{1}{k_2^{\frac{1}{2}}}
\Big)
+\frac{Q(r_{\perp,3})Q(r_{\perp,4})}
     {k_3^{\frac{3}{2}}k_4^{\frac{3}{2}}}
\Big(
 \frac{1}{k_3^{\frac{1}{2}}}
+\frac{1}{k_4^{\frac{1}{2}}}
\Big)
\Big]
\nonumber\\
&=&
\frac{M^5}{H_0^5}
\frac{(k_1+k_2)^5 H_0^{\frac{7}{2}}}{k^5}
\Big[
\frac{Q(r_{\perp, 1})Q(r_{\perp,2})}
     {k_1^{\frac{3}{2}}k_2^{\frac{3}{2}}}
\Big(
 \frac{1}{k_1^{\frac{1}{2}}}
+\frac{1}{k_2^{\frac{1}{2}}}
\Big)
+\frac{Q(r_{\perp,3})Q(r_{\perp,4})}
     {k_3^{\frac{3}{2}}k_4^{\frac{3}{2}}}
\Big(
 \frac{1}{k_3^{\frac{1}{2}}}
+\frac{1}{k_4^{\frac{1}{2}}}
\Big)
\Big],
\eea
which is of order $\frac{M^5}{H_0^5}\epsilon_{\ast}^7$.
Thus the direction dependent trispectrum 
can be as large as 
that in the de Sitter inflation
for $\frac{M}{H_0}>\epsilon_{\ast}^{-\frac{7}{5}}$.
But this is never satisfied because of Eq. (\ref{sam}).
In addition,
assuming that $k=k_1+k_2+k_3+k_4$ is fixed,
the ratio of 
the trispectrum of case II, Eq. (\ref{con4_lim}), 
to that of case I, Eq. (\ref{con_lim_a}),
is given by
$\Big(\frac{H_0}{k}\Big)^{\frac{3}{2}}$,
which is suppressed.

On the other hand,
for the planar modes,
the trispectrum 
in the same limit
is given by
\bea
T_c(\bk_1,\bk_2,\bk_3,\bk_4)
&\to &
-\frac{3
H_0^{8} \eta_0^5}
    {5 (k_1k_2k_3k_4)
}
\nonumber\\
&\times&
\frac{e^{-\pi (q_{1,1}+q_{1,2})}\sin \big(2\psi_1+2\psi_2\big)
     +e^{-\pi (q_{1,3}+q_{1,4})}\sin \big(2\psi_3+2\psi_4\big)}
     {(1-e^{-2\pi q_{1,1}})(1-e^{-2\pi q_{1,2}})
      (1-e^{-2\pi q_{1,3}})(1-e^{-2\pi q_{1,4}})}.
\label{t_p5}
\eea
The ratio to the case of the de Sitter inflation
is given by
\bea
\frac{M^5}{H_0^5}
\frac{(k_1+k_2)^5}{k^5}
\frac{e^{-\pi (q_{1,1}+q_{1,2})}\sin (2\psi_1+2\psi_2)
+e^{-\pi( q_{1,3}+q_{1,4})}\sin (2\psi_3+2\psi_4)}
{(1-e^{-2\pi q_{1,1}})
(1-e^{-2\pi q_{1,2}})
(1-e^{-2\pi q_{1,3}})
(1-e^{-2\pi q_{1,4}})},
\eea
which is greater than unity
since $\frac{M}{H_0}\gg 1$.
This bound is consistent with Eq. (\ref{sam})
for $1<\frac{M}{H_0}<\epsilon_{\ast}^{-1}$.
The ratio of Eq. (\ref{t_p1})
to Eq. (\ref{t_p5})
is given by 
\bea
\frac{e^{-\pi q_{1,1}}\sin (2\psi_1)}
{e^{-\pi (q_{1,1}+q_{1,2})} \sin (2\psi_1+2\psi_2)
+ e^{-\pi (q_{1,3}+q_{1,4})} \sin (2\psi_3+2\psi_4)},
\eea
which is of order unity,
in contrast to the one
for the nonplanar high-momentum modes.

Similar results can be obtained 
for the case of $k_1+k_4\to k_2+k_3$.

\subsubsection{Case III}

In the squeezed limit of 
$k_1= k_2\gg k_{12}$ and $k_3= k_4\gg k_{12}$,
the leading order contribution to the trispectrum is given by
\bea
T_c(\bk_1,\bk_2,\bk_3,\bk_4)
&\to &
\frac{3
     H_0^8 \eta_0^5}
    {5(k_1^2 k_3^2)}
{\rm Im}
\Big\{
\big(A^+_{{\bf k}_1}-A^-_{{\bf k}_1}\big)^{\ast}
\big(A^+_{{\bf k}_2}-A^-_{{\bf k}_2}\big)^{\ast}
\big(A^+_{{\bf k}_3}-A^-_{{\bf k}_3}\big)^{\ast}
\big(A^+_{{\bf k}_4}-A^-_{{\bf k}_4}\big)^{\ast}
\nonumber\\
&\times&
\Big(
A_{\bk_1}^{+}
A_{\bk_2}^{-}
+
A_{\bk_1}^{-}
A_{\bk_2}^{+}
\Big)
\Big(
A_{\bk_3}^{+}
A_{\bk_4}^{-}
+
A_{\bk_3}^{-}
A_{\bk_4}^{+}
\Big)^{\ast}
\Big\}
\nonumber\\
&\simeq &
\frac{3
H_0^{\frac{23}{2}} \eta_0^5}
    {20(k_1^{\frac{7}{2}} k_3^{\frac{7}{2}})
}
 \Big(\frac{1}{k_1^{\frac{1}{2}}}
    +\frac{1}{k_3^{\frac{1}{2}}}\Big)
\big(Q(r_{\perp,1})+Q(r_{\perp,2})\big)
\big(Q(r_{\perp,3})+Q(r_{\perp,4})\big).
\label{lin2}
\eea
The ratio of Eq. (\ref{lin2}) to 
that in the case of the 
de Sitter inflation, Eq. (\ref{tds1}),
is given by 
\bea
&&
\frac{H_0^{\frac{7}{2}}\eta_0^5 (k_1+k_2)^5}
     {k_1^{\frac{3}{2}}k_3^{\frac{3}{2}}}
\Big(
 \frac{1}{k_1^{\frac{1}{2}}}
+\frac{1}{k_3^{\frac{1}{2}}}
\Big)
\big(
Q(r_{\perp,1})
+Q(r_{\perp,2})
\big)
\big(
Q(r_{\perp,3})
+Q(r_{\perp,4})
\big)
\nonumber\\
&=&
\frac{M^5}{H_0^5}
\frac{H_0^{\frac{7}{2}}(k_1+k_2)^5}
     {k_1^{\frac{3}{2}}k_3^{\frac{3}{2}} k^5}
\Big(
 \frac{1}{k_1^{\frac{1}{2}}}
+\frac{1}{k_3^{\frac{1}{2}}}
\Big)
\big(
Q(r_{\perp,1})
+Q(r_{\perp,2})
\big)
\big(
Q(r_{\perp,3})
+Q(r_{\perp,4})
\big),
\eea
which is of order $\frac{M^5}{H_0^5}\epsilon_{\ast}^7$.
Thus the direction dependent trispectrum 
can be as large as 
that in the de Sitter inflation
for $\frac{M}{H_0}>\epsilon_{\ast}^{-\frac{7}{5}}$.
But this is never satisfied because of Eq. (\ref{sam}).
In addition,
assuming that $k=k_1+k_2+k_3+k_4$ is fixed,
the ratio of 
the trispectrum of case II, Eq. (\ref{con4_lim}),
to that of case III, Eq. (\ref{lin2}),
is of order $O(1)$,
and the ratio to case I, (\ref{con_lim_a}),
is suppressed by $\epsilon_\ast^3$.

On the other hand,
for the planar modes, 
\bea
T_c(\bk_1,\bk_2,\bk_3,\bk_4)
&\to &
-
\frac{3
H_0^8 \eta_0^5}
    {5(k_1^2 k_3^2)
}
\frac{1}
     {(1-e^{-\pi q_{1,1}})(1-e^{-\pi q_{1,2}})
      (1-e^{-\pi q_{1,3}})(1-e^{-\pi q_{1,4}})}
\nonumber\\
&\times&
\Big[ 
     e^{-\pi(q_{1,1}+q_{1,3})}\sin (2\psi_1+2\psi_3)
     +e^{-\pi(q_{1,1}+q_{1,4})}\sin (2\psi_1+2\psi_4)
\nonumber\\
&    +&e^{-\pi(q_{1,2}+q_{1,3})}\sin (2\psi_2+2\psi_3)
     +e^{-\pi(q_{1,2}+q_{1,4})}\sin (2\psi_2+2\psi_4)
\Big].
\label{t_p3}
\eea
The ratio to the case of the de Sitter inflation
is given by
\bea
&&\frac{M^5}{H_0^5}
\frac{(k_1+k_3)^5}{k^5}
\frac{1}
     {(1-e^{-\pi q_{1,1}})(1-e^{-\pi q_{1,2}})
      (1-e^{-\pi q_{1,3}})(1-e^{-\pi q_{1,4}})}
\nonumber\\
&\times&
\Big[ 
     e^{-\pi(q_{1,1}+q_{1,3})}\sin (2\psi_1+2\psi_3)
     +e^{-\pi(q_{1,1}+q_{1,4})}\sin (2\psi_1+2\psi_4)
\nonumber\\
&    +&e^{-\pi(q_{1,2}+q_{1,3})}\sin (2\psi_2+2\psi_3)
     +e^{-\pi(q_{1,2}+q_{1,4})}\sin (2\psi_2+2\psi_4)
\Big],
\eea
which is greater than unity
since $\frac{M}{H_0}\gg 1$.
This bound is consistent with Eq. (\ref{sam})
for $1<\frac{M}{H_0}<\epsilon_{\ast}^{-1}$.
The ratio of Eq. (\ref{t_p5})
to Eq. (\ref{t_p3})
is given by 
\bea
&&\Big[ e^{-\pi (q_{1,1}+q_{1,2})} \sin (2\psi_1+2\psi_2)
+ e^{-\pi (q_{1,3}+q_{1,4})} \sin (2\psi_3+2\psi_4)\Big]
\nonumber\\
&\times&
\Big[
 e^{-\pi (q_{1,1}+q_{1,3})}\sin (2\psi_1+2\psi_3)
+e^{-\pi (q_{1,1}+q_{1,4})}\sin (2\psi_1+2\psi_4)
\nonumber\\
&+&e^{-\pi (q_{1,2}+q_{1,3})}\sin (2\psi_2+2\psi_3)
+e^{-\pi (q_{1,2}+q_{1,4})}\sin (2\psi_2+2\psi_4)
\Big]^{-1}
\eea
which is of order unity,
in contrast to the one
for the nonplanar high-momentum modes.

Similar results can be obtained 
for the case of $k_1= k_4\gg k_{14}$ and $k_2= k_3\gg k_{14}$.

\subsubsection{Features of the trispectra from the contact interaction}

We summarize 
the features of the trispectra
from the contact interaction diagram
and compare them with those in the de Sitter inflation.
In our background
the trispectrum in the limiting case I
(for example, for $k_1\to k_2+k_3+k_4$)
is greater
than those in the limiting case II (
for example, for $k_1+k_2\to k_3+k_4$)
and case III
(for example, for $k_1=k_2\gg k_{12}$ and $k_3=k_4\gg k_{12}$),
by a factor of $\big(\frac{k}{H_0}\big)^{\frac{3}{2}}$,
where $k \gg H_0$ is the typical size of the
momentum vectors.

We then summarize the direction dependence
of the trispectra. 
Though we focus on the nonplanar high-momentum modes,
the essence is the same 
for the planar modes.
In the limiting case I, 
for example, for $k_1\to k_2+k_3+k_4$,
the trispectrum
is proportional to $Q(r_{\perp,1})$
and hence 
depends on the direction of the momentum
vector ${\bf k}_1$.
In the limiting case II,
for example, for $k_1+k_2\to k_3+k_4$,
it is proportional to 
a combination of
$Q(r_{\perp,1})Q(r_{\perp,2})$
and 
$Q(r_{\perp,3})Q(r_{\perp,4})$,
which depends on the directions of all the momenta.
In the limiting case III,
for example,
for $k_1=k_2\gg k_{12}$ and $k_3=k_4\gg k_{12}$,
the trispectrum 
is proportional to
$\big(Q(r_{\perp,1})+Q(r_{\perp,2})\big)
\big(Q(r_{\perp,3})+Q(r_{\perp,4})\big)$,
which depends on the directions of all the external momenta.
Thus the trispectra 
are direction dependent,
and the dependence is significantly
different in various limiting cases.

We finally compare the amplitudes of
the trispectra from a contact interaction diagram 
with those in the de Sitter inflation.
The amplitude of the trispectra 
for the planar modes
in all the limiting cases I, II and III
can be greater than those in the de Sitter inflation.
For the nonplanar high-momentum modes,
the amplitude of the trispectrum 
in the limiting case I
can be as large as that in the de Sitter inflation,
while those in the other two limiting cases
cannot be as large as those
in the de Sitter inflation.
Thus,
combined with the direction dependence,
the trispectrum in the limiting case I
for the nonplanar high-momentum modes,
as well as those in all the limiting cases
for the planar modes,
would be particularly interesting to 
distinguish models.

\subsection{Contribution from the scalar exchange interaction}

The other leading order contribution to
the trispectrum
is given by the scalar exchange interaction.
The trispectrum is given by
\bea
&&\langle \phi(t,{\bf k}_1)\phi(t,\bk_2)
\phi(t,\bk_3)\phi(t,\bk_4)
\rangle
\nonumber\\
&=&
-
\int_{t_0}^{t}dt_2
\int_{t_0}^{t_2}dt_1
\Big\langle
\big[H^{(3)}_I (t_1),
\big[H^{(3)}_I (t_2),
\phi^I(t,\bk_1)\phi^I(t,\bk_2)
\phi^I(t,\bk_3)\phi^I(t,\bk_4)\big]
\big]
\Big\rangle,
\eea
where the interaction Hamiltonian is given by
Eq. (\ref{min}).
The trispectrum in the $t\to 0$ limit takes the following form
\bea
\langle
\phi(0,{\bf k}_1)\phi(0,\bk_2)\phi(0,\bk_3)\phi(0,\bk_4)
\rangle
=(2\pi)^{-3}\frac{q_3^2}{M^4}
T_s(\bk_1,\bk_2,\bk_3,\bk_4)
\delta(\bk_1+\bk_2+\bk_3+\bk_4).
\eea
After computations, 
we obtain
\bea
T_s(\bk_1,\bk_2,\bk_3,\bk_4)
&=&
 \frac{9
      H_0^8}
     {4(k_1k_2k_3k_4)
       }
\nonumber\\
\nonumber\\
&\times&
{\rm Re}
\Big\{
k_{12}
\big(A^+_{{\bf k}_1}-A^-_{{\bf k}_1}\big)
\big(A^+_{{\bf k}_2}-A^-_{{\bf k}_2}\big)
\big(A^+_{{\bf k}_3}-A^-_{{\bf k}_3}\big)^{\ast}
\big(A^+_{{\bf k}_4}-A^-_{{\bf k}_4}\big)^{\ast}
\nonumber\\
&\times&
\Big[
-A^+_{\bk_3}A^+_{\bk_4}A^+_{\bk_{12}}
 \cI(k_3,k_4,k_{12})
+A^-_{\bk_3}A^-_{\bk_4}A^-_{\bk_{12}}
 \cI(-k_3,-k_4,-k_{12})
\nonumber\\
&+&
\Big( A^+_{\bk_3}A^+_{\bk_{4}}A^-_{\bk_{12}}
 \cI(k_3,k_4,-k_{12})
+ {\rm 2 \,\,perms}\Big)
\nonumber\\
&-&
\Big( A^+_{\bk_3}A^-_{\bk_4}A^-_{\bk_{12}}
 \cI(k_3,-k_4,-k_{12})
+ {\rm 2 \,\,perms}\Big)
\Big]
\nonumber\\
&\times&
\Big[
-\big(A^+_{\bk_1}A^+_{\bk_2}A^+_{\bk_{12}}\big)^{\ast}
 \cI(-k_1,-k_2,-k_{12})
+\big(A^-_{\bk_1}A^-_{\bk_2}A^-_{\bk_{12}}\big)^{\ast}
 \cI(k_1,k_2,k_{12})
\nonumber\\
&+&
\Big(\big( A^+_{\bk_1}A^+_{\bk_{2}}A^-_{\bk_{12}}\big)^{\ast}
 \cI(-k_1,-k_2,k_{12})
+ {\rm 2 \,\,perms}\Big)
\nonumber\\
&-&
\Big(\big(A^+_{\bk_1}A^-_{\bk_2}A^-_{\bk_{12}}\big)^{\ast}
 \cI(-k_1,k_2,k_{12})
+ {\rm 2 \,\,perms}\Big)
\Big]
\nonumber\\
&+&
k_{14}
\big(A^+_{{\bf k}_2}-A^-_{{\bf k}_2}\big)
\big(A^+_{{\bf k}_3}-A^-_{{\bf k}_3}\big)
\big(A^+_{{\bf k}_4}-A^-_{{\bf k}_4}\big)^{\ast}
\big(A^+_{{\bf k}_1}-A^-_{{\bf k}_1}\big)^{\ast}
\nonumber\\
&\times&
\Big[
-A^+_{\bk_4}A^+_{\bk_1}A^+_{\bk_{14}}
 \cI(k_4,k_1,k_{14})
+A^-_{\bk_4}A^-_{\bk_1}A^-_{\bk_{14}}
 \cI(-k_4,-k_1,-k_{14})
\nonumber\\
&+&
\Big( A^+_{\bk_4}A^+_{\bk_{1}}A^-_{\bk_{14}}
 \cI(k_3,k_4,-k_{14})
+ {\rm 2 \,\,perms}\Big)
\nonumber\\
&-&
\Big( A^+_{\bk_4}A^-_{\bk_1}A^-_{\bk_{14}}
 \cI(k_4,-k_1,-k_{14})
+ {\rm 2 \,\,perms}\Big)
\Big]
\nonumber\\
&\times&
\Big[
-\big(A^+_{\bk_2}A^+_{\bk_3}A^+_{\bk_{14}}\big)^{\ast}
 \cI(-k_2,-k_3,-k_{14})
+\big(A^-_{\bk_2}A^-_{\bk_3}A^-_{\bk_{14}}\big)^{\ast}
 \cI(k_2,k_3,k_{14})
\nonumber\\
&+&
\Big(\big( A^+_{\bk_2}A^+_{\bk_{3}}A^-_{\bk_{14}}\big)^{\ast}
 \cI(-k_2,-k_3,k_{14})
+ {\rm 2 \,\,perms}\Big)
\nonumber\\
&-&
\Big(\big(A^+_{\bk_2}A^-_{\bk_3}A^-_{\bk_{14}}\big)^{\ast}
 \cI(-k_2,k_3,k_{14})
+ {\rm 2 \,\,perms}\Big)
\Big]
\Big\}.
\eea
Thus the trispectrum 
from the scalar exchange interaction depends on
$k_{12}=|\bk_1+\bk_2|$
and 
$k_{14}=|\bk_1+\bk_4|$
as well as
on $k_i$ ($i=1,2,3,4$)
because of the internal line in the diagram.
Since $|\cI(p_1,p_2,p_3)|$
is increasing
in the limit of $p_1+p_2+p_3\to 0$,
the bispectrum
has the maximum amplitude.

We focus on limits 
$k_1+k_2\to k_{12}$ and $k_3+k_4\to k_{12}$,
and
$k_1+k_4\to k_{14}$ and $k_2+k_3\to k_{14}$ (case I).
We also investigate the limiting cases of 
$k_1=k_2\gg k_{12}$ and $k_3=k_4\gg k_{12}$,
and 
$k_1=k_4\gg k_{14}$ and $k_2=k_3\gg k_{14}$
(case II).

\subsubsection{Case I}

In the limit of $k_1+k_2\to k_{12}$ and $k_3+k_4\to k_{12}$,
the leading order contribution to the
trispectrum is given by
\bea
T_s(\bk_1,\bk_2,\bk_3,\bk_4)
&\to&
 \frac{
H_0^8 k_{12}\eta_0^6}
     {4(k_1k_2k_3k_4)
}
\nonumber\\
&\times&
{\rm Re}
\Big\{ \big(A^+_{{\bf k}_1}-A^-_{{\bf k}_1}\big)
\big(A^+_{{\bf k}_2}-A^-_{{\bf k}_2}\big)
\big(A^+_{{\bf k}_3}-A^-_{{\bf k}_3}\big)^{\ast}
\big(A^+_{{\bf k}_4}-A^-_{{\bf k}_4}\big)^{\ast}
\nonumber\\
&\times&
\Big[
 A^+_{{\bf k}_3} A^+_{{\bf k}_4} A^-_{{\bf k}_{12}}
-A^-_{{\bf k}_3} A^-_{{\bf k}_4} A^+_{{\bf k}_{12}}
\Big]
\Big[
\Big(
A^+_{{\bf k}_1} A^+_{{\bf k}_2} A^-_{{\bf k}_{12}}
\Big)^{\ast}
-
\Big(A^-_{{\bf k}_1} A^-_{{\bf k}_2} A^+_{{\bf k}_{12}}
\Big)^{\ast}
\Big]\Big\}.
\eea
For the trispectrum of
the nonplanar high-momentum modes,
we obtain
\bea
&&
{\rm Re}
\Big\{\big(A^+_{{\bf k}_1}-A^-_{{\bf k}_1}\big)
\big(A^+_{{\bf k}_2}-A^-_{{\bf k}_2}\big)
\big(A^+_{{\bf k}_3}-A^-_{{\bf k}_3}\big)^{\ast}
\big(A^+_{{\bf k}_4}-A^-_{{\bf k}_4}\big)^{\ast}
\nonumber\\
&\times&
\Big[
 A^+_{{\bf k}_3} A^+_{{\bf k}_4} A^-_{{\bf k}_{12}}
-A^-_{{\bf k}_3} A^-_{{\bf k}_4} A^+_{{\bf k}_{12}}
\Big]
\Big[
\Big(
A^+_{{\bf k}_1} A^+_{{\bf k}_2} A^-_{{\bf k}_{12}}
\Big)^{\ast}
-
\Big(A^-_{{\bf k}_1} A^-_{{\bf k}_2} A^+_{{\bf k}_{12}}
\Big)^{\ast}
\Big]
\Big\}
\simeq 
\frac{Q(r_{\perp, 12})^2 H_0^3}{4k_{12}^3}.
\eea
Thus
\bea
T_s(\bk_1,\bk_2,\bk_3,\bk_4)
\to
 \frac{
H_0^{11} \eta_0^6 
        Q(r_{\perp,12})^2}
     {16(k_1k_2k_3k_4)
 k_{12}^2}.
\label{sca_lim}
\eea
The ratio to the case of the de Sitter inflation,
Eq. (\ref{tds2}),
is given by 
\bea
k_{12}^3 H_0^3\eta_0^6 
Q(r_{\perp,12})^2
=\frac{M^6}{H_0^6}
\frac{k_{12}^3H_0^3}{k^6}
Q(r_{\perp,12})^2,
\eea
which is of order $\frac{M^6}{H_0^6}\epsilon_{\ast}^6$.
Thus the direction dependent trispectrum 
can be as large as 
that in the de Sitter inflation
for $\frac{M}{H_0}>\epsilon_{\ast}^{-1}$.
This is marginally consistent with Eq. (\ref{sam}).

On the other hand, 
the trispectrum 
in the same limit is given by 
\bea
T_s(\bk_1,\bk_2,\bk_3,\bk_4)
\to
\frac{
H_0^8 k_{12}\eta_0^6}
     {4(k_1k_2k_3k_4)}
\frac{e^{-2\pi q_{1,12}}}
 {(1-e^{-2\pi q_{1,1}})
  (1-e^{-2\pi q_{1,2}})
  (1-e^{-2\pi q_{1,3}})
  (1-e^{-2\pi q_{1,4}})
  (1-e^{-2\pi q_{1,12}})}.
\nonumber\\
&&
\label{t_p2}
\eea
The ratio to the case of the de Sitter inflation
is given by
\bea
&&\frac{M^6}{H_0^6}
\frac{k_{12}^6}{k^6}
\frac{e^{-2\pi q_{1,1}}}
     {(1-e^{-\pi q_{1,1}})(1-e^{-\pi q_{1,2}})
      (1-e^{-\pi q_{1,3}})(1-e^{-\pi q_{1,4}})},
\eea
which is greater than unity
since $\frac{M}{H_0}\gg 1$.
This bound is consistent with Eq. (\ref{sam})
for $1<\frac{M}{H_0}<\epsilon_{\ast}^{-1}$.

Similar results can be obtained
for $k_1+k_4\to k_{14}$ and $k_2+k_3\to k_{14}$.

\subsubsection{Case II}

In the limit of 
$k_1= k_2\gg k_{12}$ and $k_3= k_4\gg k_{12}$,
the leading order contribution 
from the nonplanar high momentum modes is given by
\bea
T_s(\bk_1,\bk_2,\bk_3,\bk_4)
&\to &
\frac{
 H_0^{11} \eta_0^6 k_{12}}
    {16 (k_1^{\frac{7}{2}} k_3^{\frac{7}{2}})
    }
\big(Q(r_{\perp,1})
    +Q(r_{\perp,2})
\big)
\big(
     Q(r_{\perp,3})
    +Q(r_{\perp,4})
\big).
\label{koh}
\eea
which is of the same order as Eq. (\ref{sca_lim}).
The ratio to the case of the de Sitter inflation, Eq. (\ref{tds2}),
is given by 
\bea
&&
(k_1 k_3)^{\frac{3}{2}}
H_0^3\eta_0^6 
\big(Q(r_{\perp,1})+Q(r_{\perp,2})\big)
\big(Q(r_{\perp,3})+Q(r_{\perp,4})\big)
\nonumber\\
&=&
\frac{M^6}{H_0^6}
\frac{(k_1 k_3)^{\frac{3}{2}}
H_0^3}{k^6}
\big(Q(r_{\perp,1})+Q(r_{\perp,2})\big)
\big(Q(r_{\perp,3})+Q(r_{\perp,4})\big),
\eea
which is of order $\frac{M^6}{H_0^6}\epsilon_{\ast}^6$.
Thus the direction dependent trispectrum 
can be as large as 
that in the de Sitter inflation
for $\frac{M}{H_0}>\epsilon_{\ast}^{-1}$.
This is marginally consistent with Eq. (\ref{sam}).
In addition,
the trispectrum in case II, Eq. (\ref{koh}),
is suppressed to that in case I, Eq. (\ref{sca_lim}),
by the ratio of $\frac{k_{12}}{k}\ll 1$.
Thus in contrast to the case of the bispectrum,
the trispectrum in case I becomes
greater than that in case II.

On the other hand,
for the planar modes,
\bea
T_s(\bk_1,\bk_2,\bk_3,\bk_4)
&\to &
\frac{
H_0^{8} \eta_0^6}
    {4 (k_1^2 k_3^2)
}
\frac{k_{12}}
     {(1-e^{-\pi q_{1,1}})(1-e^{-\pi q_{1,2}})
      (1-e^{-\pi q_{1,3}})(1-e^{-\pi q_{1,4}})
       (1-e^{-\pi q_{1,12}})}
\nonumber\\
&\times&
\Big\{
e^{-\pi (q_{1,1}+q_{1,3})}
\cos\big(2\psi_1-2\psi_3)
+e^{-\pi (q_{1,1}+q_{1,4})}
\cos\big(2\psi_1-2\psi_4)
\nonumber\\
&+&e^{-\pi (q_{1,2}+q_{1,3})}
\cos\big(2\psi_2-2\psi_3)
+e^{-\pi (q_{1,2}+q_{1,4})}
\cos\big(2\psi_2-2\psi_4)
\Big\}.
\label{pink}
\eea
The ratio to the case of the de Sitter inflation
is given by
\bea
&&\frac{M^6}{H_0^6}
\frac{k_{1}^3k_3^3}{k^6}
\frac{1}
     {(1-e^{-\pi q_{1,1}})(1-e^{-\pi q_{1,2}})
      (1-e^{-\pi q_{1,3}})(1-e^{-\pi q_{1,4}})
       (1-e^{-\pi q_{1,12}})}
\nonumber\\
&\times&
\Big\{
e^{-\pi (q_{1,1}+q_{1,3})}
\cos\big(2\psi_1-2\psi_3)
+e^{-\pi (q_{1,1}+q_{1,4})}
\cos\big(2\psi_1-2\psi_4)
\nonumber\\
&+&e^{-\pi (q_{1,2}+q_{1,3})}
\cos\big(2\psi_2-2\psi_3)
+e^{-\pi (q_{1,2}+q_{1,4})}
\cos\big(2\psi_2-2\psi_4)
\Big\},
\eea
which is greater than unity
since $\frac{M}{H_0}\gg 1$.
This bound is consistent with Eq. (\ref{sam})
for $1<\frac{M}{H_0}<\epsilon_{\ast}^{-1}$.
The ratio of Eq. (\ref{t_p2})
to Eq. (\ref{pink}) is given by 
\bea
&&\frac{ke^{-2\pi q_{1,12}}}{k_{12}}
  \Big[e^{-\pi (q_{1,1}+q_{1,3})}\cos (2\psi_1-2\psi_3)
     +e^{-\pi (q_{1,1}+q_{1,4})}\cos (2\psi_1-2\psi_4)
\nonumber\\
     &+&e^{-\pi (q_{1,2}+q_{1,3})}\cos (2\psi_2-2\psi_3)
     +e^{-\pi (q_{1,2}+q_{1,4})}\cos (2\psi_2-2\psi_4)
 \Big]^{-1},
\eea
which can be larger than unity for $k_{12}\ll k$
as for the nonplanar high momentum modes.
The result is in contrast to the case of the
bispectrum from the contact diagram.

Similar results can be obtained 
for the case of $k_1= k_4\gg k_{14}$ and $k_2= k_3\gg k_{14}$.

\subsubsection{Features of the trispectra from the scalar exchange 
interaction}

We summarize 
the features of the trispectra
from the scalar exhange interaction diagram
and compare them with those in the de Sitter inflation.
In our background
the amplitude of the trispectrum in the limit of case I
(for example, for $k_1 + k_2\to k_{12}$ and $k_3+k_4\to k_{12}$)
is greater than those in the limits of 
case II 
(for example, for $k_1=k_2\gg k_{12}$ and $k_3=k_4\gg k_{12}$),
by a factor of $\frac{k}{k_{12}}$,
where $k \gg H_0$ is the typical size of the
momentum vectors.

We then summarize the direction dependence
of the trispectra.
Though we focus on 
the nonplanar high-momentum modes,
the essence is the same
for the planar modes.
In the limiting case I, 
for example, 
for $k_1 + k_2\to k_{12}$ and $k_3+k_4\to k_{12}$),
the trispectrum
is proportional to $Q(r_{\perp,{12}})^2$
and hence 
depends on the direction of the internal momentum
vector ${\bf k}_{12}$.
In the limiting case II,
for example, for 
$k_1=k_2\gg k_{12}$ and $k_3=k_4\gg k_{12}$,
it is proportional to 
$\big(Q(r_{\perp,1})+Q(r_{\perp,2})\big)
\big(Q(r_{\perp,3})+Q(r_{\perp,4})\big)$,
which depends on the directions of all the momenta.
Thus the trispectra 
are direction dependent,
and the dependence is significantly
different in various limiting cases.

We finally compare the amplitudes of
the trispectra from the scalar exchange interaction diagram 
with those in the de Sitter inflation.
The amplitudes of the trispectra from
the planar modes
in all the limiting cases I and II
can be greater than those in the de Sitter inflation.
For the nonplanar high-momentum modes,
the amplitudes of the trispectra
in both the limiting cases I and II
can be at least comparable to those in the de Sitter inflation.
Thus, 
combined with the direction dependence,
the trispectra in
all the limiting cases 
for the nonplanar high-momentum modes,
as well as those for the planar modes,
would be interesting
to distinguish models.

\section{Summary}

We have investigated the higher order spectra of a scalar field
induced by the higher order time-derivative interactions,
which becomes dominant above the energy scale $M$,
in the universe with a primordial anisotropy.
Along the lines of our recent works \cite{km,km2},
a change in the vacuum causes
new features in the higher order spectra.
We have evaluated the leading order contributions
to the bispectra and trispectra.

The leading order contribution
to the bispectra is given by
the three-point interaction.
The mixing of the negative frequency mode
gives peaks of the bispectra
in the limits of $k_1\to k_2+k_3$ and its 
permutation,
where $k_i=|{\bf k}_i|$ 
are the magnitudes of the momentum vectors of a triangle.
An imporant feature of the bispectra 
is that 
in the limit of $k_3\to k_1+ k_2$
they are suppressed to that
in the limit of $k_1=k_2\gg k_3$
by the ratio $\frac{k_3}{k}$,
in fixing the total momentum $k=k_1+k_2+k_3$.
For the nonplanar high-momentum modes,
for various shapes of the triangle
the ratio
to the case of the de Sitter inflation
is given by $\frac{M^3}{H_0^3}\epsilon^4_{\ast}$,
which is larger than unity 
for $\frac{M}{H_0}>\epsilon_{\ast}^{-\frac{4}{3}}$,
where $\epsilon_{\ast}$ is the adiabaticity parameter
evaluated at the instance of the matching.
However, this bound is not consistent with 
$\frac{M}{H_0}<\epsilon_{\ast}^{-1}$,
which is obtained from the requirement 
that the time when the momentum of a mode becomes smaller than
the energy scale of the derivative interactions
becomes later than the time when the de Sitter mode function starts to become valid.
For the planar modes,
the bispectra
are always larger than 
in the de Sitter inflation
by a factor $\big(\frac{M}{H_0}\big)^3<\epsilon_{\ast}^{-3}$.
In summary,
for the nonplanar high-momentum modes
the bispectra 
cannot be as large as that of 
de Sitter inflation.
On the other hand,
for the planar modes,
they can be greater than those
in the de Sitter inflation.
The bispectra are direction dependent,
and the dependence is significantly different
in various limiting cases.
Thus the bispectra for the planar modes 
would be particularly interesting to distinguish models.

We have also 
investigated the trispectra
in the same spacetime background.
The leading order
contribution to the trispectra
is obtained from two kinds of diagrams:
the 
contact interaction
and
the scalar exchange interaction diagrams
(see, e.g., \cite{tri} for the Feynman diagrams).
We have first considered
the nonplanar high-momentum modes.
The trispectrum from
the contact interaction
is amplified
in the limit of $k_1\to k_2+k_3+k_4$
(and in its permutation),
compared to
the case of the de Sitter inflation.
The ratio of the trispectrum in the limit
of $k_3\to k_1+ k_2+k_3$
in the initially anisotropic universe
to that in the isotropic one
is given by $\frac{M^5}{H_0^5}\epsilon^4_{\ast}$,
which is larger than unity
for $\frac{M}{H_0}>\epsilon_{\ast}^{-\frac{4}{5}}$.
This bound can be consistent with 
$\frac{M}{H_0}<\epsilon_{\ast}^{-1}$.
Thus 
the trispectrum in this limit
can be enhanced 
for smaller energy scales $M$
than in the case of the bispectrum. 
In the other important limits,
its ratio becomes 
of order $\frac{M^5}{H_0^5}\epsilon^7_{\ast}$,
which is larger than unity 
for $\frac{M}{H_0}>\epsilon_{\ast}^{-\frac{7}{5}}$.
However, this bound cannot be consistent with 
$\frac{M}{H_0}<\epsilon_{\ast}^{-1}$.
Thus it is hard
to enhance these shapes of the trispectra.
For the planar modes,
the trispectra from the contact interactions
are always larger than
in the de Sitter inflation
by a factor $\big(\frac{M}{H_0}\big)^5<\epsilon_{\ast}^{-5}$.
In summary,
for the nonplanar high-momentum modes,
the trispectra 
can be as large as that of de Sitter inflation
only for the limit of $k_1\to k_2+k_3+k_4$
(and also for its permutations).
On the other hand,
for the planar modes,
these can be greater than those
in the de Sitter inflation.
The trispectra are direction dependent,
and its dependence is significantly different
in various limiting cases.
Thus for the nonplanar high-momentum modes,
the trispectra in 
this limiting case
would be particularly interesting to distinguish models.
Also, as for the bispectra, 
in all the limiting cases
the planar modes would be important.

In the case of the trispectrum from
the scalar exhange interactions,
the amplitude in the 
the limit of $k_1+k_2\to |\bk_1+\bk_2|$ and
$k_3+k_4\to |\bk_1+\bk_2|$
(and also for $k_1+k_4\to |\bk_1+\bk_4|$
and $k_2+k_3\to |\bk_1+\bk_4|$)
can be more important
than that 
in the limit of $k_1=k_2\gg |\bk_1+\bk_2|$ and
$k_3=k_4\gg |\bk_1+\bk_2|$,
by the ratio $\frac{k}{k_{12}}$.
This result is in contrast to the case of the bispectra
and would give an unique signature
from an initially anisotropic universe.
For the nonplanar high-momentum modes, 
the ratio of the trispectra in the initially
anisotropic universe
to those in the case of the de Sitter inflation
is given by $\frac{M^6}{H_0^6}\epsilon^6_{\ast}$,
which can be more important 
for $\frac{M}{H_0}>\epsilon_{\ast}^{-1}$.
This bound can be marginally consistent with 
$\frac{M}{H_0}<\epsilon_{\ast}^{-1}$.
For the planar modes,
the trispectra from the contact interactions
are always larger than 
in the de Sitter inflation
by a factor $\big(\frac{M}{H_0}\big)^6<\epsilon_{\ast}^{-6}$.
In summary,
for the nonplanar high-momentum modes
the trispectra
can be as large as that 
in de Sitter inflation
in all the limiting cases.
Also, for the planar modes,
they can be greater than those
in the de Sitter inflation.
The trispectra are direction dependent,
and its dependence is significantly different
in various limiting cases.
Thus for the nonplanar high-momentum modes,
the trispectra in all the limiting cases
would be particularly interesting to distinguish models.
Also, as for the bispectra, 
in all the limiting cases
the planar modes would be important.


Before closing this article,
we would like to mention 
the directions of our future studies.
One of the most important directions is
to formulate the
nonlinear cosmological
 perturbations theory in the anisotropic universe
and 
evaluate the higher order spectra.
Since
in the anisotropic cosmological background,
even at the level of the linear perturbations,
there is a significant mixture of the different types of 
perturbations in terms of three-dimensional symmetry,
the same effects may also become important
during the isotropization,
as well as in the change of the vacuum state.
Another important direction is
to cure the initial curvature singularity
in other types of initially anisotropic spacetime
which is singular,
and to present a successful quantization scheme of the 
perturbations.
They may be achieved by the matter fields,
which could significantly backreact on the 
spacetime geometry 
and regularize the initial singularity.

\section*{Acknowledgements}
We thank H-c. Kim for fruitful discussions.
This work was supported by Yukawa fellowship and 
by Grant-in-Aid for Young Scientists (B) of JSPS Research,
under Contract No. 24740162.


\appendix

\section{Higher order spectra
of a scalar field in the de Sitter universe}

In this appendix,
as a reference,
we discuss the higher spectra
of a scalar field
in the de Sitter universe.
induced by the derivative interactions 
given in the main text \eqref{min}.

\subsection{Bispectrum}

The bispectrum from the contact interaction 
in the de Sitter inflation is given by 
\bea
\label{bds}
B(\bk_1,\bk_2,\bk_3)
= -\frac{3
H_0^5 
}
        {(k_1 k_2 k_3)(k_1+k_2+k_3)^3
}.
\eea

\subsection{Trispectrum}

\subsubsection{Contribution from the contact interaction}

The trispectrum from the contact interaction
in the de Sitter inflation is given by 
\bea
\label{tds1}
T_c(\bk_1,\bk_2,\bk_3,\bk_4)
&=&
 \frac{72
H_0^8}
        {(k_1 k_2 k_3k_4)(k_1+k_2+k_3+k_4)^5 
}.
\eea

\subsubsection{Contribution from the scalar exchange interaction diagram}

The trispectrum from the scalar exchange interaction diagram
in the de Sitter inflation
is given by 
\bea
\label{tds2}
T_s(\bk_1,\bk_2,\bk_3,\bk_4)
&=&
\frac{9
        H_0^8}
     {2(k_1k_2k_3k_4)
}
\Big\{
\frac{k_{12}}
     {(k_1+k_2+k_{12})^3(k_3+k_4+k_{12})^3}
+
\frac{k_{14}}
     {(k_2+k_3+k_{14})^3(k_1+k_4+k_{14})^3}
\Big\}.
\eea


\end{document}